# On-chip spectropolarimetry by fingerprinting with random surface arrays of nanoparticles

*Yiting Chen*[1], Fei Ding[1], Victor Coello[2], and Sergey I. Bozhevolnyi*[1]*

[1]Centre for Nano Optics, University of Southern Denmark, Campusvej 55, DK-5230 Odense M, Denmark

[2]Unidad Monterrey, Centro de Investigación Cientfica y de Educacion Superior de Ensenada (CICESE), Alianza Centro 504, Apodaca, NL, 66629, Mexico

**ABSTRACT:** Optical metasurfaces revolutionized the approach to moulding the propagation of light by enabling simultaneous control of the light phase, momentum, amplitude and polarization. Thus, instantaneous spectropolarimetry became possible by conducting parallel intensity measurements of differently diffracted optical beams. Various implementations of this very important functionality have one feature in common - the determination of wavelength utilizes dispersion of the diffraction angle, requiring tracking the diffracted beams in space. Realization of on-chip spectropolarimetry calls thereby for conceptually different approaches. In this work, we demonstrate that random nanoparticle arrays on metal surfaces, enabling strong multiple scattering of surface plasmon polaritons (SPPs), produce upon illumination complicated SPP scattered patterns, whose angular spectra are uniquely determined by the polarization and wavelength of light, representing thereby spectropolarimetric fingerprints. Using μm-sized circular arrays of randomly distributed nm-sized gold nanoparticles (density ~ 75 μm$^{-2}$) fabricated on gold films, we measure angular distributions of scattered SPP waves using the leakage radiation microscopy and find that the angular SPP spectra obtained for normally incident light beams different in wavelength and/or polarization are distinctly different. Our approach allows one to realize on-chip spectropolarimetry by fingerprinting using surface nanostructures fabricated with simple one-step electron-beam lithography.

Light-matter interactions using metal nanostructures have opened a way to the manipulation of light beyond the diffraction limit based on the propagation and localization of surface plasmon polaritions (SPPs), hybrid waves involving electromagnetic fields in dielectrics and collective free-electron oscillations in metals, with the corresponding field termed plasmonics.[1,2] With unique characteristics featuring extremely broad bandwidth, strong field enhancements and subwavelength confinement of electromagnetic energy, plasmonics has not only given rise to various applications, such as nonlinear signal conversion,[3-5] light harvesting,[6-8] lasering[9-11] and sensing,[12-15] but also offered the opportunity to bridge the gap between photonics and electronics with miniature devices.[16,17] Furthermore, specially designed metal surface nanostructures, termed plasmonic metasurfaces[18], have also demonstrated their ability to function as ultrathin planar optical devices and to control the amplitude, phase and polarization of light by arranging plasmonic resonant antennas of specific size, shape and distribution.[18-22] One of the most important and recent advances in this direction concerns the development of spectropolarimeters,[23-25] which enable simultaneous measurements of the spectrum and state of polarization (SOP) of light and constitute a powerful analytic tool with capabilities far exceeding those of separate polarimeters and spectrometers. Metasurfaces employed in these spectropolarimeters operate analogously to SOP-sensitive blazed diffraction gratings by directing different polarizations into spatially separated diffraction orders, whose diffraction angles are determined by the incident light wavelength.

While on-chip metasurface-only polarimeters allowing for the simultaneous determination of all Stokes parameters have also been demonstrated,[26,27] on-chip spectropolarimetry with any of conventional metasurfaces seems to represent mission impossible because of principal difficulties in encoding two independent information (polarimetric and spectroscopic) channels into one-dimensional parameter (polar angle of in-plane propagation) space. Considering this issue from the viewpoint of surface wave excitation, the problem is that the conversion from incident propagating waves to surface modes requires phase matching along the surface wave propagation, which is typically achieved by using periodic structures, so that changing the wavelength influences only the diffraction efficiency and not the propagation direction. Realization of the on-chip spectropolarimetry calls thereby for conceptually different (from gradient metasurfaces[23-27]) approaches that would allow to realize the aforementioned encoding.

Random (disordered) arrays of surface nanoparticles (NPs) represent a striking alternative to periodic arrays exploited in metasurfaces. It is worth mentioning that the term "plasmonic metasurfaces" was first introduced when considering SPP scattering phenomena by random and periodic nanostructures, because these could conveniently be described applying the concept of SPP effective index inside nanostructured areas.[28] Random metal nanostructures have been found to exhibit versatile abilities of controlling light, such as subwavelength light localization[29] and second-harmonic enhancement through multiple SPP scattering,[30] white light generation,[31] SPP routing by scattering-free channels,[32,33] and multiple wavefront shaping.[34]

Underlying physical mechanisms involved in the aforementioned phenomena have often in common strong multiple SPP scattering and localization initiated either by external illumination[29,30] or by launching an SPP beam towards a scattering NP array.[32,33] In general, localization of light happens because of interference in recurrent multiple scattering in random media, resulting in inhibition of light propagation when the scattering mean free path decreases below the light wavelength.[35] Since two-dimensional waves can be strongly localized with any degree of disorder (in the absence of dissipation), the only condition for the strong SPP localization is that the SPP propagation length should be much larger than the SPP localization length.[29,32] In practice, this implies that, due to the exponential divergence of the localization length with the scattering mean-free path, the latter should be similar to or smaller than the SPP wavelength ($l \leq \lambda_{SPP}$). In the previous experiments on SPP scattering by random gold NP arrays fabricated using electron-beam lithography (EBL), typical values of the SPP scattering mean-free path at near-infrared wavelengths were $l \sim 250$ nm,[33,36] indicating that the regime of strong multiple SPP scattering and localization was realized in these random nanostructures.

In this work, we demonstrate that random nanoparticle arrays on metal surfaces, enabling strong SPP multiple scattering, produce upon illumination complicated SPP scattered patterns, whose angular spectra are uniquely determined by the polarization and wavelength of light, representing thereby spectropolarimetric fingerprints. We fabricate dense μm-sized circular arrays of randomly distributed nm-sized gold NPs, which are known to ensure the regime of strong SPP multiple scattering at near-infrared wavelengths,[32, 33, 36] and observe SPP excitation and scattering produced by these arrays upon illumination with normally incident light beams of different wavelengths and/or SOPs. Using the leakage radiation microscopy (LRM) in the image and Fourier planes,[33] we find that the angular distributions of scattered SPP waves form very complicated irregular patterns that are distinctly different for different wavelengths and/or SOPs of the incident light. We relate this well-pronounced effect to far-field interference of SPP waves originating from uncorrelated (and spatially localized) SPP sources generated upon illumination by strong SPP multiple scattering.[29] We further demonstrate that these visually different SPP scattering spectra can be quantitatively discriminated using the correlation analysis. Numerical simulations of angular SPP scattering spectra conducted using the previously developed self-consistent approach[37,38] are found in good qualitative agreement with experimental data, emphasizing the importance of realizing the regime of strong multiple SPP scattering. The obtained results validate thereby the proposed concept of on-chip spectropolarimetry by fingerprinting based on SPP excitation and scattering by random surface nanostructures. Although this approach requires calibration of fabricated nanostructures (by measuring the angular spectra for various SOPs and wavelengths), i.e., registering fingerprints, the design and fabrication procedure is exceedingly simple, involving only generation of NP random coordinates that are subsequently used in the straightforward one-step EBL or nanoimprint lithography fabrication. Note that the comparison of an unknown SPP scattering spectrum (i.e., obtained with an unknown SOP and wavelength) to the calibrated spectra from the database (i.e., the fingerprint identification), can greatly be facilitated using the wavelet-based data processing.[39]

## RESULTS

**LRM imaging of SPP scattering.** The concept proposed for on-chip spectropolarimetry is based on the hypothesis that outgoing SPP waves, produced by strongly scattering random NP arrays upon illumination, form very complicated irregular patterns (Figure 1a) that are distinctly different for incident light beams different in wavelength and/or SOPs. In order to demonstrate the validity of this approach we fabricated dense (~ 75 $\mu m^{-2}$) circularly shaped (with radii $R$ from 1 to 5 μm) random arrays of gold 60-nm-wide and 70-nm-high NPs atop a 70-nm-thick gold film supported by a 170-μm-thick silica glass substrate (Figure 1b). The fabrication procedure involved deposition of a gold film and one-step EBL fabrication of random NP arrays (see the Methods in Supporting Information). Although gold NPs were attempted to produce with diameters of 50 nm, the resulting NPs were often slightly larger by up to 20 nm due to proximity effects causing aggregation as observed in the fabricated samples with scanning electron microscope (SEM) images (Figure 1c,d). The choice of NP size and density was motivated by the experimental results of our previous studies of SPP scattering, localization and waveguiding in the EBL-fabricated random NP arrays.[32,33,36] These parameters might not be optimum for the considered purpose but have already proven to be sufficient for ensuring the regime of strong multiple SPP scattering and localization.

LRM imaging of the SPP excitation and scattering by the fabricated arrays was conducted using a tunable Ti-Sapphire laser with the output range of 720–1000 nm, whose output beam was horizontally polarized while quarter-wave and half-wave plates were used along with two linear polarizers to control the SOP of the beam incident on the random arrays (Figure S1). The direct and Fourier images of SPP scattering were captured by two infrared (IR) charge coupled devices (CCDs) in both image (Figure 2a) and Fourier (Figure 2b) planes with the aperture stop introduced in the Fourier plane center to remove the directly transmitted laser beam. We observed in the Fourier plane that the angular distributions of scattered SPP waves form very complicated irregular patterns. Furthermore, as the random array diameter was increased, the angular SPP scattering pattern became progressively more complicated (Figure 3a-3h). We relate these features to the occurrence (upon illumination of random NP arrays) of strong multiple SPP scattering and localization resulting in the formation of uncorrelated and spatially localized (within the SPP wavelength) SPP sources, whose strengths and phases are determined by the incident light SOP and wavelength. The strongest influence on the SPP far-field scattering pattern and its fine structure is expected to originate from the SPP sources located along the outer perimeter of random NP arrays. Since the smallest angular width of interference maxima in the far-field is determined by the SPP diffraction and estimated to be ~ $\lambda_{SPP}/R$, the total number of SPP rays (maxima in the angular SPP scattering pattern) can be evaluated simply as $N \sim 2\pi R/\lambda_{SPP}$. It turned out that this formula describes fairly well the experimental results obtained for random arrays with $R$ = 1, 2, 3, and 5 μm, predicting $N$ ~ 8, 16, 24 and 39 with the numbers of observed (Figure 3i-3l) maxima being $N$ = 9, 21, 25, and 32. In the remaining part of the paper, we mainly focus on the results obtained for the 3-μm-radius random NP array that is already large enough to demonstrate excellent performance as elaborated also when discussing the results obtained.

The Fourier image (Figure 2b), representing a circle with the radius corresponding to the SPP wave-vector magnitude, offers a convenient way for qualitatively characterizing the angular SPP scattering spectra by averaging CCD pixel signals near the SPP circle along the same azimuthal angle $\theta$ (Materials and methods). We observed that the angular SPP scattering spectra although being very irregular were stable with respect to the beam size (once that was large enough) but changed drastically whenever the SOP (Figure 2c) and/or light wavelength was changed. These unique SPP scattering spectra represent thereby spectropolarimetric fingerprints of the incident light beam and can conveniently be presented using multi-line coloured bar encoding (Figure 2d) that facilitates visual comparison of different SPP scattering spectra.

**Correlation of angular spectra for different SOPs and wavelengths.** The SPP scattering spectra obtained for different SOPs or light wavelengths were quantitatively compared by calculating the correlation coefficients $C_{ij}$ (Materials and methods). Considering four main SOPs, we found that the spectra corresponding to the orthogonal (linear or circular) polarizations are practically uncorrelated: $C_{ij} < 0.2$ (Figure 4). At the same time, the correlation coefficients between the spectra (obtained with differently sized random areas and at different wavelengths) corresponding to linear and circular polarizations were found to be dispersed around but close to 0.5 with the upper bound decreasing gradually for larger random areas (Table S1). This difference in the correlation of spectra corresponding to different SOPs can be related to the circumstance that any circular polarization has one electrical field component identical to that of any linear polarization (Supplementary Section 1). Taking into account that the (close to) identical angular spectra should have the unitary (or very close to) correlation coefficient, we conclude that the polarimetric fingerprints of the main SOPs are substantially different. This conclusion can also be supported by direct visual observations when comparing the corresponding rows in the angular SPP scattering spectra (Figure 4).

After establishing the uniqueness of the angular spectra for substantially different SOPs, we investigated the influence of the angle $\alpha$ between two linear polarizations by recording the angular SPP scattering spectra with differently-sized random NP areas for different angles $\alpha$ with respect to the horizontal polarization. It is seen that angular dependences of the correlation coefficient become close to the analytical expression: $C(\alpha)=cos^2\alpha$, once the random area size becomes sufficiently large (Figure 5a). The correlation coefficient decreases thereby with increasing $\alpha$ in the same way as the polarizer transmission when increasing the angle between the incident polarization and the polarizer orientation. Considering highly irregular features of the angular SPP scattering spectra used to calculate the correlation coefficient, this seems surprising but can be argued for in a straightforward manner as in the previous case of correlation between linear and circular polarizations (Supplementary Section 1). Although direct comparison of two angular spectra obtained with two different linear polarizations is not very sensitive to small angles between these polarizations, one can definitely increase the accuracy of the polarization angle determination by comparing the angular spectrum of an unknown polarization with all angular spectra recorded for different polarization angles and fitting the corresponding correlation coefficients to $cos^2\alpha$ dependence.

We further investigated correlation between angular SPP scattering spectra obtained for different illumination wavelengths, observing a rapid decrease in the correlation coefficient for increasing differences in wavelengths (Figure 5b). Following the same arguments regarding the occurrence of uncorrelated and spatially localized SPP sources (producing angular scattering spectra), one can reason that these sources are formed by multiple SPP scattering and interference within random cavities, whose quality factors are of the order of $L_{SPP}/\lambda$ (neglecting the SPP dispersion) where $L_{SPP}$ is the SPP propagation length. The wavelength selectivity can then be estimated as $\delta\lambda \sim \lambda^2/L_{SPP}$, which results in the wavelength sensitivity $\delta\lambda \sim 27$ nm, when using the following parameters:[16] $L_{SPP} \sim 24$ μm at $\lambda \sim 800$ nm. The sensitivity evaluated in this way is consistent with the experimental results (Figure 5b), representing essentially a ballpark estimation. It is clear that, also in this case, one can definitely increase the accuracy of the wavelength identification by comparing the angular spectrum for an unknown wavelength with all angular spectra recorded for different wavelengths.

## DISCUSSION

The challenge of determining the wavelength and SOP of an optical beam with planar (on-chip) configurations has been approached in our work by making use of strong multiple SPP scattering in random NP arrays. In order to elucidate the importance of strong multiple SPP scattering, we conducted numerical simulations using the previously developed self-consistent approach to describe the multiple light scattering by NPs represented within the framework of point-dipole approximation, with the Green dyadic being approximated by analytic expressions available for the near- and far-field regions.[37,38] Although there are certain limitations to the accuracy of this approach,[40] the simulations were generally found to accurately reflect the main physical features in various scattering phenomena.[38] Likewise in this case, we have found in our simulations of SPP scattering spectra features similar to the experimentally observed ones: influence of the scattering area size, difference in the correlation coefficients when comparing linear and circularly polarized incident beams, deterioration in the correlation for misaligned linear polarizations and different wavelengths (Supplementary Section 2). Additionally, our simulations highlighted the importance of realization of strong multiple SPP scattering for differentiating right and left circular polarizations. It is clear that, when individual NPs are isotropic in the scattering plane, weak (single) scattering cannot produce different angular SPP scattering spectra, whereas strong multiple scattering couples many NPs that are sufficiently away from each other, introducing thereby chirality in the scattering response (Figure S3 and Table S2). Overall, the obtained results validate the proposed concept of on-chip spectropolarimetry by fingerprinting based on SPP excitation and scattering by random surface nanostructures.

The operation principle introduced in our work requires careful calibration of a fabricated NP array by measuring the angular SPP scattering spectra for various SOPs and wavelengths in a procedure analogous to the registration of fingerprints. The creation of a database of spectropolarimetric fingerprints is a bit tedious but straightforward procedure, and the database search for a match can be accomplished with the correlation analysis and facilitated using well-developed wavelet-based data processing techniques.[39] On the positive side, the design and fabrication procedure of a (strongly) scattering NP array is exceedingly simple, involving only generation of random coordinates for (sufficiently large and densely packed[32,33,36]) NPs so that these coordinates can subsequently be used in the straightforward one-step EBL or nanoimprint lithography fabrication. Note that measurements of the angular SPP scattering spectra to be used for the database do not impose stringent requirements on the incident laser beam size (it should be just large enough) and angular alignment (normal incidence with one-degree accuracy is easy to realize and sufficient for reproducible measurements). Moreover, gold nanostructures and films are known to be stable in normal environment and suitable for long-term applications as noticed in the experiments using gold nanostructures and films for SPP localization[29-33], gap-plasmon based gradient metasurfaces[20,25,28] and V-groove plasmonic waveguides[11]. Otherwise, it is also very simple to coat these structures with a thin layer of $SiO_2$ for better protection. Furthermore, we believe that this approach can also be applied to identification of optical fields with different orbital angular momenta[24] using the same principle, although in this case the alignment of the optical beam to be examined becomes very important because of inhomogeneous spatial distributions of the incident field phase and amplitudes. Taking into account explosively increasing interest in spin-orbit photonics,[41] scattering of optical vortex beams by random nanostructures deserves special studies, promising new possibilities for spin-orbit optical interactions.[42] Overall, the proposed fingerprinting principle opens, in our opinion, a way to on-chip realization of general recognition of complex optical field patterns with possible applications not only in the spectropolarimetry but also within optical wavefront sensing, optical beam tracing and positioning (including beam autofocusing) as well as environmental sensing, circumventing the usual requirements of elaborated design and precise nanofabrication.

## Methods

**Fabrication**. Commercially available (18×18 mm$^2$) 170-μm-thick silica glass substrates were cleaned in an ultrasonic bath with distilled water, acetone and isopropyl alcohol, during 5 minutes consecutively, and covered with a 3-nm-thin layer of Ti by means of thermal evaporation in order to improve the adhesion to gold. After depositing a 70-nm-thick gold film, a substrate was spin coated with poly-methyl-methacrylate (PMMA) (~ 350 nm thick) serving as a resist layer for the EBL. Disordered patterns were designed using a simple MATLAB program based on "random" function to generate random coordinates of points with the density of 75 μm$^{-2}$ within a circle of a given radius (ranging from 1 to 5 μm), discarding in the process of generation the points that were too close to each other (i.e., requiring the minimum distance to be larger than 50 nm). Thus created file of random coordinates was used in the EBL system to conduct the PMMA film patterning. The final fabrication step involved the deposition of a 70-nm-thick gold film and lift-off in acetone.

**LRM imaging and data processing**. Fourier images of SPP scattering (Figure 2b) were treated using a simple MATLAB program for computing the angular SPP intensity distributions. First, we manually found the origin ($k = 0$) and the radius ($k = N_{SPP}$ – the SPP effective index) corresponding to a given Fourier image. The angular SPP scattering spectra were then computed with the step of one degree in the azimuthal angle $\theta$ (Figure 2b) by averaging the signals from three neighbour pixels along a given angle. Since, in the experiment, we only needed to rotate the wave plates and tune the laser to change the SOP and wavelength of the incident light, the origin and radius in the Fourier images were typically the same for the same structure (the wavelength dispersion of SPP effective index is negligibly small in the wavelength range used). This circumstance allowed us to simplify the image treatment procedure significantly. Once the SPP scattering spectra were determined, resulting in ordered arrays of 360 values (different for different SOPs and wavelengths of the incident light), the correlations between different spectra, $I_{m(n)}(\theta)$, were calculated using standard formulae:[43] $corr(I_m, I_n) = cov(I_m, I_n)/\sigma_m\sigma_n$, with $cov(I_m, I_n) = \langle[I_m(\theta) - \langle I_m\rangle] \cdot [I_n(\theta) - \langle I_n\rangle]\rangle$, $\langle I_{m(n)}\rangle = \frac{1}{360}\sum_{\theta=1}^{360} I_{m(n)}(\theta)$, and $\sigma_{m(n)} = \sqrt{\langle(I_{m(n)}(\theta) - \langle I_{m(n)}\rangle)^2\rangle}$.

ASSOCIATED CONTENT

**Supporting Information**

Figures S1 to S4, Table S1 to S3, calculations and simulations of angular SPP scattering spectra, reference list (PDF).

AUTHOR INFORMATION

**Corresponding Authors** *E-mail: yic@mci.sdu.dk; seib@mci.sdu.dk

**ORCID**

Yiting Chen: 0000-0002-4003-4796

Sergey I. Bozhevolnyi: 0000-0002-0393-4859

**Notes** The authors declare no competing financial interest.

ACKNOWLEDGEMENT

The authors gratefully acknowledge financial support from the European Research council, Grant 341054 (PLAQNAP) and the University of Southern Denmark (SDU 2020).

**Figures**

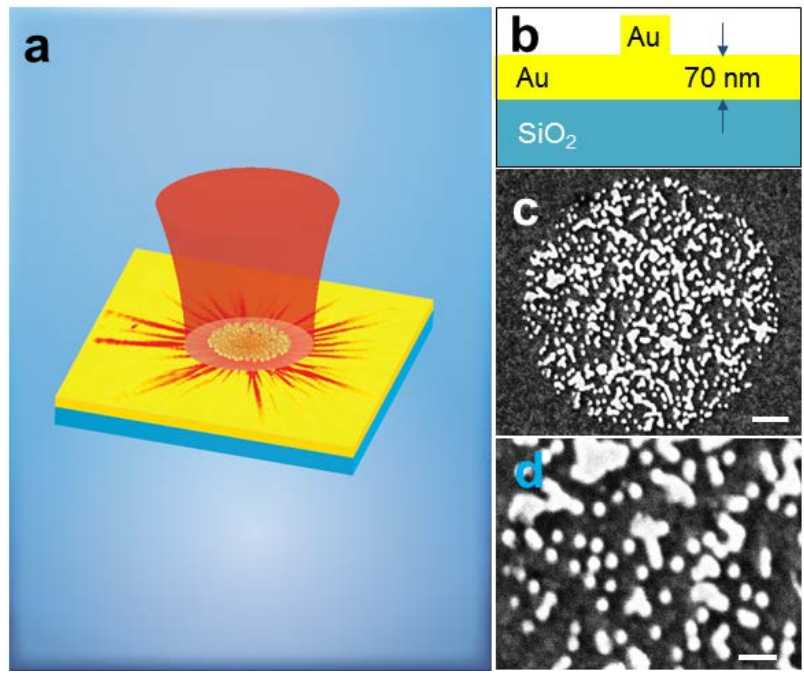

**Figure 1.** (a) Schematic of SPP excitation and scattering occurring upon illumination of random NP arrays at normal incidence. (b) Sketch of an individual gold NP atop a 70-µm-thick gold film deposited on a silica wafer. (c) Scanning electron microscope (SEM) image of a fabricated random NP array with the radius of 2 µm. Scale bar, 0.5 µm. (d) Enlarged part of the SEM image shown in c. Scale bar, 200 nm.

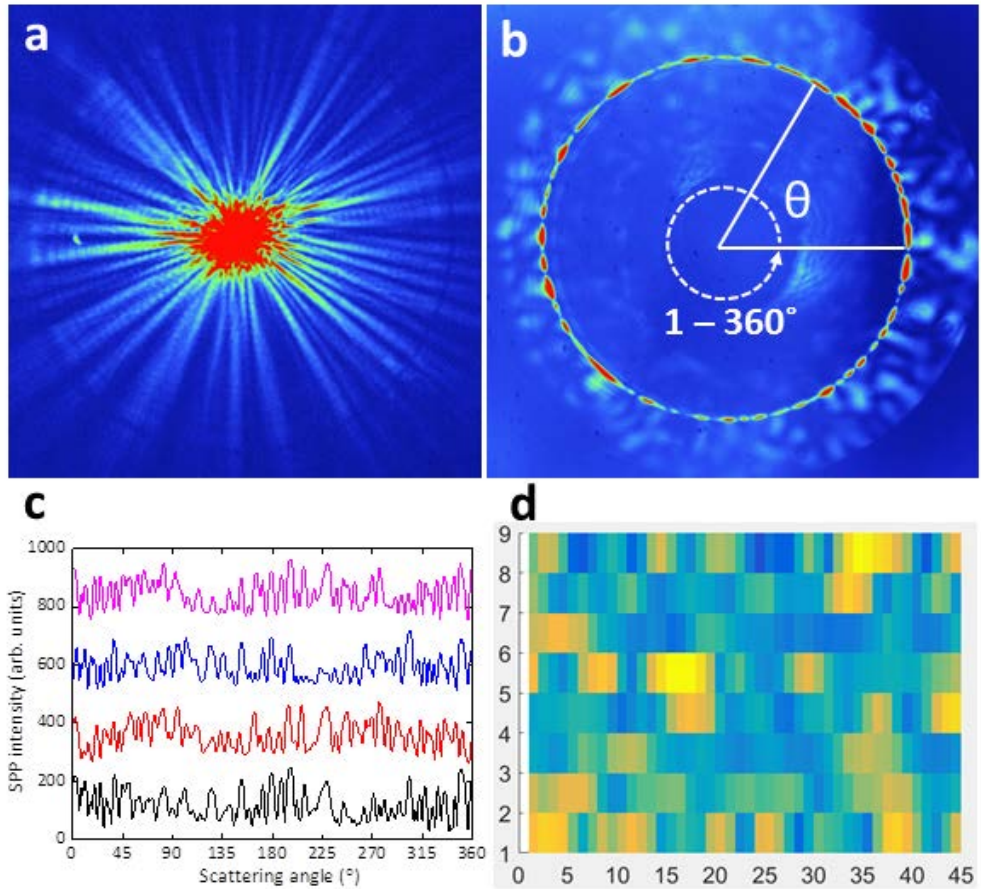

**Figure 2.** (a) LRM image of the SPP excitation and scattering in the direct (sample surface) plane obtained with a random 5-μm-radius array being illuminated by a linearly polarized (in the horizontal direction) laser beam at the wavelength of 800 nm. (b) The corresponding LRM image in the Fourier plane introducing the azimuthal angle $\theta$ that denotes the SPP scattering direction. The bright circle is formed by SPP waves scattered in different directions with its radius being related to the SPP effective index. (c) Angular SPP scattering spectra recorded with four different SOPs of incident light at 800 nm, which are, from bottom to top, vertical polarization (black), horizontal polarization (red), left-circular polarization (blue) and right-circular polarization (magenta). (d) The SPP scattering spectrum obtained with the vertical polarization (shown also in c) and represented, for convenience in visual comparison of SPP spectra, using 8 coloured stripes (barcodes) displaying sequentially, from left to right and from top to bottom, 45°-wide angular distributions.

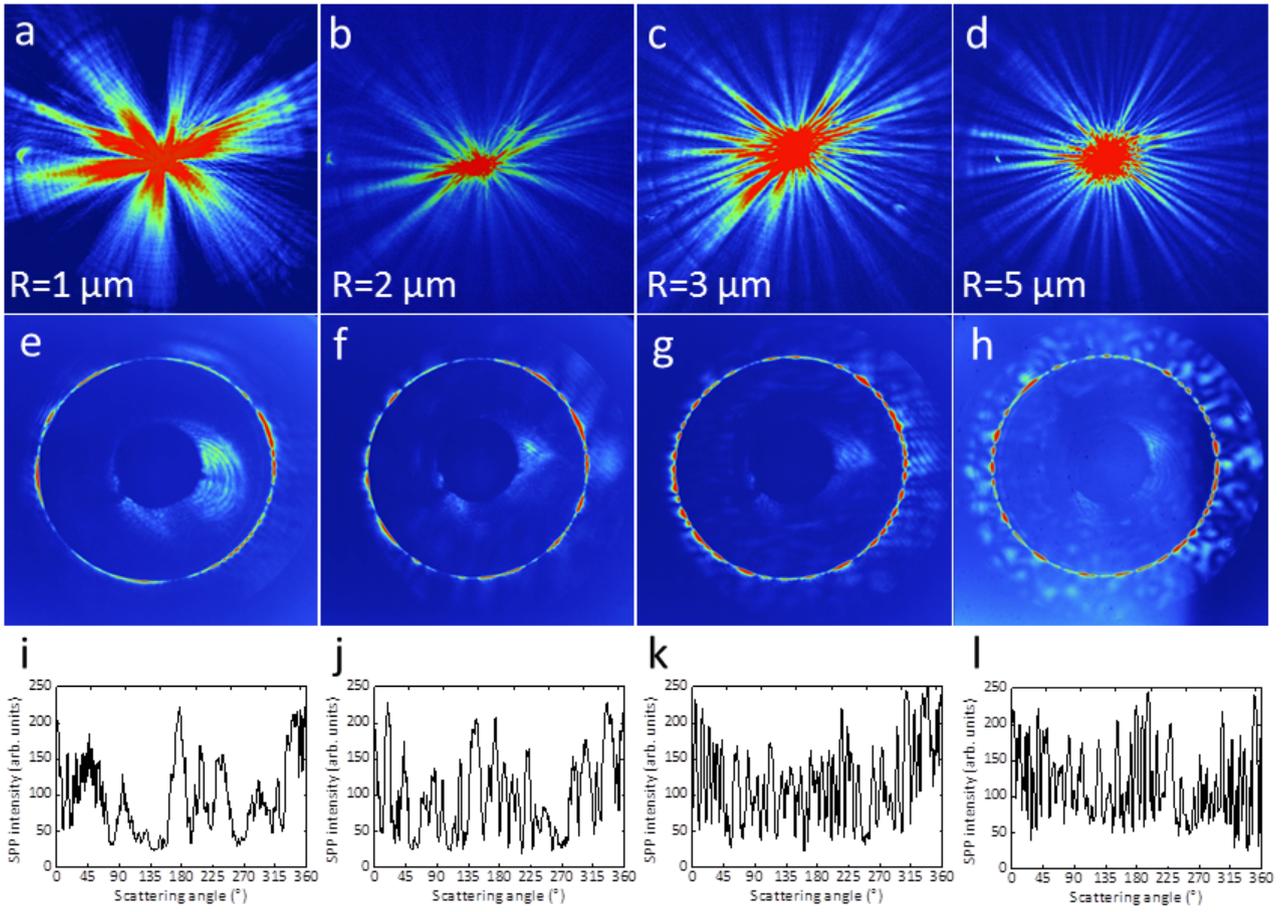

**Figure 3.** LRM characterization of differently-sized random areas. a – d, LRM images of scattered SPP waves obtained in the (direct) image plane with linearly polarized (in the horizontal direction) light at the illumination wavelength of 800 nm that is normally incident on differently-sized random NP arrays with radii $R$ = a, 1, b, 2, c, 3 and d, 5 μm, respectively. e – h, The corresponding images obtained in the Fourier plane with the respective random areas and with the transmitted beam spot being blocked. i – l, The corresponding angular SPP intensity distributions retrieved from the respective Fourier images. The numbers of distinct maxima in these distributions are identified as 9, 21, 25, and 32, respectively, by using a criterion that a local maximum should be situated between two local minima with signal drops of more than 50% with respect to the maximum value.

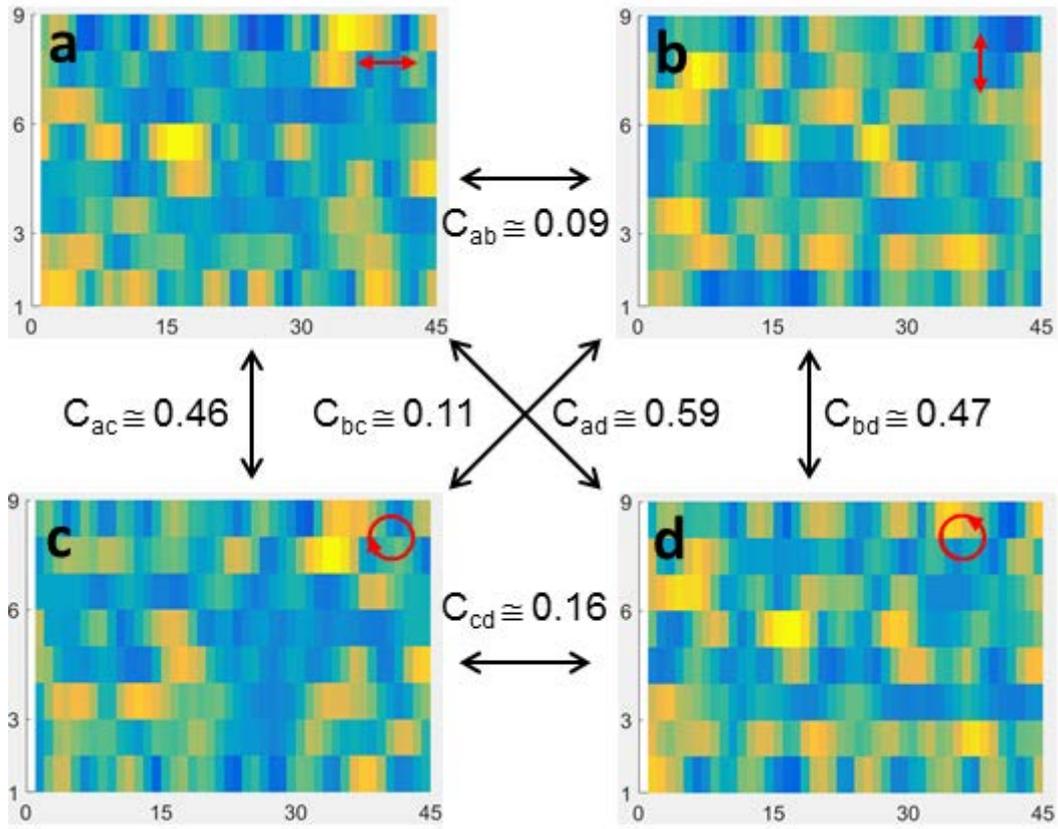

**Figure 4.** Multi-line barcodes representing angular SPP scattering spectra obtained with a random 5-μm-radius array being illuminated at the wavelength of 800 nm with (a) horizontal linear, (b) vertical linear, (c) left-circular and (d) right-circular polarizations. $C_{mn}$ indicate the corresponding correlation coefficients between the spectra displayed.

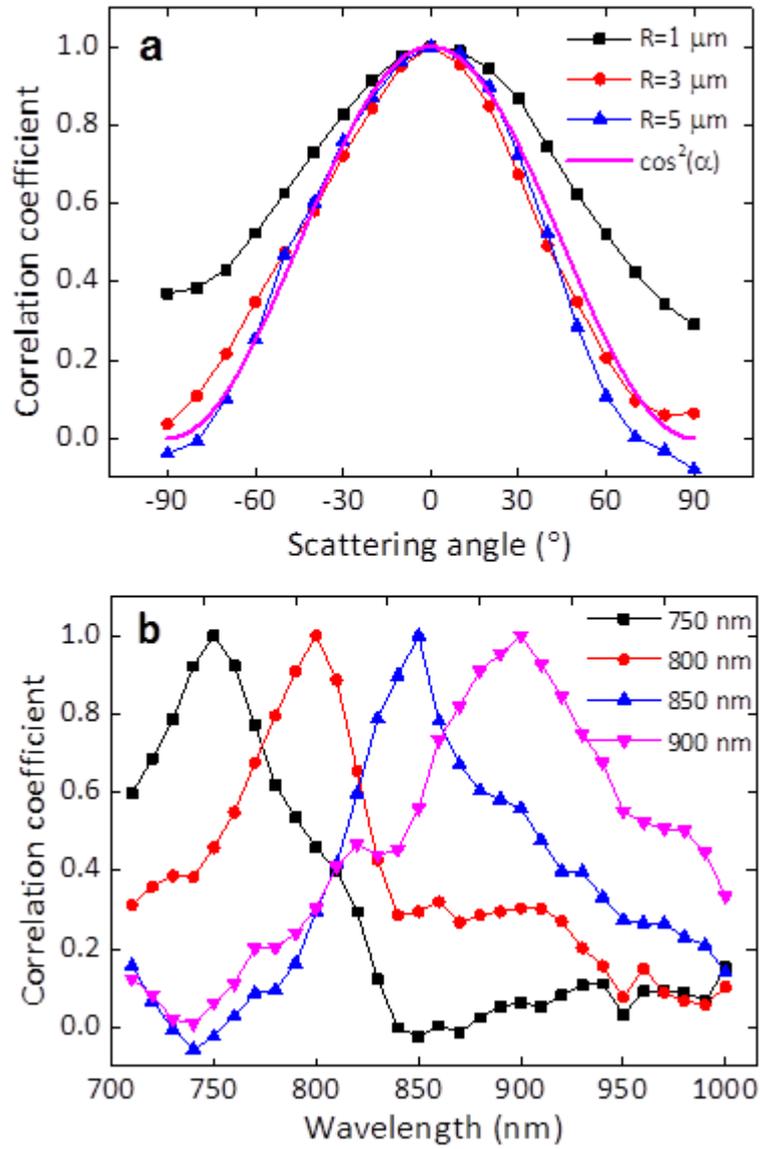

**Figure 5.** (a) Correlation between SPP scattering spectra obtained for different linear polarizations (at 800 nm) with differently sized random NP arrays as a function of the angle between the polarizations. The qualitatively expected dependence, $\cos^2\alpha$, is also indicated for comparison. (b) Correlation between SPP scattering spectra obtained for different incident light wavelengths with linear horizontal polarization as a function of the wavelength, when using different reference wavelengths, including 750, 800, 850 and 900 nm.